\documentstyle[emulateapj,apjfonts,psfig]{article}

\lefthead{S.~Schmoll et al.}  
\righthead{Fairall 9}

\received{Date}
\revised{Date}
\accepted{Date}

\journalid{vol}{date}
\articleid{1}{4}
\paperid{id}

\cpright{AAS}{1999}
\ccc{x}

\begin{document}


\title{Constraining the Spin of the Black Hole in Fairall 9 with Suzaku}

\author{S. Schmoll\altaffilmark{1},J. M. Miller\altaffilmark{1},
M. Volonteri\altaffilmark{1},E . Cackett\altaffilmark{1}$^{,}$\altaffilmark{2},
C. S. Reynolds\altaffilmark{3}, A. C. Fabian\altaffilmark{4},\\
L. W. Brenneman\altaffilmark{5}, G. Miniutti\altaffilmark{6}$^{,}$\altaffilmark{7},$^{,}$\altaffilmark{8}, L. C. Gallo\altaffilmark{9}}
\altaffiltext{1}{Department of Astronomy and Astrophysics, University
  of Michigan, 500 Church Street, Ann Arbor, Michigan, 48109, USA}
\altaffiltext{2}{Chandra Postdoctoral Fellow}
\altaffiltext{3}{Department of Astronomy, The University of Maryland,
College Park, Maryland, 20742}
\altaffiltext{4}{Institute of Astronomy, University of Cambridge,
Madingley Road,Cambridge CB3 OHA,UK}
\altaffiltext{5}{NPP Postdoctoral Fellow (ORAU); NASA GSFC, mail code
  662, Greenbelt MD 20771}
\altaffiltext{6}{Max-Plack-Institut f\"{u}r extraterrestrische Physik,
  Postfach 1312, 85741 Garching, Germany}
\altaffiltext{7}{Laboratoire APC, UMR 7164, 10 rue A. Domon et
  L. Duquet, 75205 Paris, FR}
\altaffiltext{8}{LAEX, Centro de Astrobiologia (CSIC-INTA), LAEFF,
  P.O. Box 78, E-28691, Villanueva de la Canada, Madrid ES}
\altaffiltext{9}{Department of Astronomy and Physics, Saint
Mary's University, 923 Robie Street, Halifax NS B3H 3CS}


\authoremail{schmoll@umich.edu}

\label{firstpage}

\begin{abstract}
We report on the results of spectral fits made to data obtained from a
168~ksec {\it Suzaku} observation of the Seyfert-1 galaxy Fairall 9.
The source is clearly detected out to 30~keV.  The observed spectrum
is fairly simple; it is well-described by a power-law with a soft
excess and disk reflection.  A broad iron line is detected, and easily
separated from distinct narrow components owing to the resolution of
the CCDs in the X-ray Imaging Spectrometer (XIS).  The broad line is
revealed to be asymmetric, consistent with a disk origin.  We fit the
XIS and Hard X-ray Detector (HXD) spectra with
relativistically-blurred disk reflection models.  With the assumption
that the inner disk extends to the innermost stable circular orbit,
the best-fit model implies a black hole spin parameter of $a =
0.60\pm0.07$ and excludes extremal values at a high level of
confidence.  We discuss this result in the context of Seyfert
observations and models of the cosmic distribution of black hole spin.
\end{abstract}

\section{Introduction}
The centers of active galactic nuclei (AGN) contain a supermassive
black hole (SMBH) that acts as the central engine by actively
accreting matter.  A topic of intensive study in AGN research is the
radio-loud/radio-quiet dichotomy in the the Eddington
ratio/radio-luminosity plane (Sikora, Stawarz, \& Lasota 2007).  There
is an apparent morphological distinction with radio-loud galaxies only
hosted by ellipticals, while radio-quiet galaxies are found to be
either ellipticals or spirals (Wilson \& Colbert 2005).  Theoretical
work on this problem has suggested that supermassive black hole spin
may be a major contributor to determining if a galaxy is radio-loud or
radio-quiet.  Such models invoke the Blandford-Znajek effect to
extract spin energy from the hole in order to power the radio jets
(Blandford \& Znajek 1977).

Astrophysical black holes have only two properties: mass and spin.  It
is likely that all black holes spin on some level; spin can be
described by a dimensionless parameter defined as $\hat{a} = cJ/GM^2$,
where J is the angular momentum of the black hole and
$0<\hat{a}<0.998$ (Thorne 1974).  Spin is a constantly evolving
parameter for a black hole because every time it accretes matter or
goes through a merger, angular momentum is transferred.

The cosmic distribution of black hole spin parameters encodes vital
aspects of black hole-galaxy co-evolution, and recent work has been
devoted to this topic.  Broadly speaking, there are two different
suggestions on what the spin distribution should look like.  Volonteri
et al. (2005) argued that the lifetime of quasars is long enough that
the innermost regions of accretion disks typically align with black
hole spins, while the direction of the angular momentum of the
accreted material is constant throughout the quasar activity.  These
models predicts that most black holes in quasars should have large
spins.  Volonteri et al. (2007) also suggest that the evolution of
black hole spins, after the quasar epoch, depends on the detailed
accretion history, which is linked to the morphology of the
host. Volonteri et al. (2007) predict that black holes in elliptical
galaxies tend to retain large spins. The distribution of spins for
black holes hosted in disk galaxies (e.g. Seyferts) is instead
predicted to be much shallower, with a tail extending to low spin
values.

Alternatively, it has been suggested that accretion always proceeds
via small (and short) uncorrelated episodes (``chaotic accretion",
King \& Pringle 2006), caused by fragmentation of the accretion disk
where it becomes self-gravitating.  This scenario implies that black
hole spins are very low ($\simeq 0.1-0.3$): accretion of randomly
oriented droplets of gas would rapidly spin down any black hole, since
counter-rotating material spins black hole down more efficiently than
co-rotating material spins them up.  To resolve this matter,
observational spin constraints are needed from as many SMBHs as
possible.

Spectroscopy of Fe K emission lines that are formed in the inner
accretion disk provide one way to constrain the spin of a black
hole (Miller 2007).  These disklines are highly skewed due to
relativistic Doppler shifts and gravitational red-shifts effects
(Laor 1991).  Especially if one assumes the accretion disk extends
to the inner-most stable circular orbit (ISCO; 

\centerline{\psfig{file=f1.ps,width=3.3in,angle=-90}}
\figcaption[h]{In the picture above, the continuum was fit with a simple
  power-law between 2-3 and 7-10 keV. There is a slight excess seen
  towards the softer X-rays which can be modeled with a 0.2 keV
  blackbody.  At high energy, some of the flux excess is consistent
  with disk reflection.  The XIS0, XIS1, XIS3, and HXD spectra are
  shown in black, red, green, and blue, respectively.}
\medskip

\noindent Bardeen et al.\ 1970), the
extremity of the line shifts at the inner disk can be used to infer
the spin of the hole.

In this paper, we report initial constraints on the spin of the SMBH
in the Seyfert-1 AGN Fairall 9 ($z=0.047$), using a relativistic disk
line.  Compared to other Seyferts where iron lines have been studied
in detail, Fairall 9 is an order of magnitude more massive and more
luminous.  Its central black hole has a mass of $2.55\pm 0.56\times
10^8$ M$_{\odot}$ based on reverberation mapping (Peterson et al.\
2004).  The Eddington fraction has been measured to be as high as 0.16
(Done \& Gierlinski 2005).  Our data reduction procedure is described
in \S2.  A description of the assumptions and various models we used
to obtain the spin of Fairall 9 follows in \S3. In \S4, we discuss our
results and their implications.

\section{Data and Reduction}
We observed Fairall 9 with the {\it Suzaku} X-ray Telescope.  The 168
ksec run started on 2007 June 7 at 3:34:52 (TT). The data were taken
using the XIS and HXD detectors, using the XIS pointing position.  The
three XIS CCD cameras (XIS0, XIS1,and XIS3) cover 0.2-12.0 keV while
the HXD/PIN covers the 10.0-60.0 keV band. The XIS data were taken in
both 3x3 and 5x5 binning models.  Exposures made using the 3x3 mode
lasted 150 ksec while the 5x5 exposures lasted approximately 18.5
ksec.  The HXD/PIN detector ran in default mode with an exposure time
of approximately 136.5 ksec. 

We reduced the version-2 processed data using the HEASOFT reduction
and analysis suite, version 6.4.  For the XIS0 and XIS3 cameras, which
are front-illuminated CCDs, we extracted source counts from a 3.1'
(180 pixel) region around the center of the source.  Background counts
were extracted from annili between 3.1' and 8.1' (180-466 pixels).
For XIS1, the background was taken from 3.6' to 8.3' (212-486 pixels).

We created XIS redistribtion matrix files (rmf) and a ancillary
response files (arf) using the HEASOFT commands ``xisrmfgen'' and
``xissimarfgen'', respectively.  The arf files were created using a
simulation that propagates photons through a model of the telescope;
we ran the simulation using the recommended 400,000 photons.  Using
these files, we grouped the 

\centerline{\psfig{file=f2.ps,width=3.3in,angle=-90}}
\figcaption[h]{The plot above shows the data/model ratio in Fe K region that
  results from a simple power-law fit to the data.  The narrow
  Gaussian peak near 6.1 keV (6.4 keV in the rest frame) is due to
  reflection from distant gas.  A broad diskline component is also
  clearly present.  The XIS0, XIS1, and XIS3 spectra are shown in
  black, red, and blue, respectively.}
\medskip
\vspace{0.1in}
\centerline{\psfig{file=f3.ps,width=3.3in,angle=-90}}
\figcaption[h]{The plot above shows the data/model ratio in the Fe K
  region, after fitting the continuum and narrow emission lines.  A
  relativistically-broadened diskline is clearly detected.  The XIS0,
  XIS1, and XIS3 spectra are shown in black, red, and blue,
  respectively.  }
\medskip

\noindent data from each of the 3x3 and 5x5 binning
modes to create one spectral file for each of the XIS detectors.

The HXD/PIN spectrum was extracted and corrected for deadtime using
the standard tools.  The deadtime fraction amounted to only a few
percent.  We then used the non X-ray background file from {\it Suzaku}
that corresponded to our observation to correct its exposure time to
match the background.  We then modeled the cosmic X-ray background
using a response file provided from {\it Suzaku} for the XIS nominal
pointing using XSPEC v.11.3.  The corrected non-X-ray background file
was added to the simulated cosmic X-ray background file; the combined
file served as the background file for analysis of the HXD/PIN
spectrum.

In our subsequent analysis of the XIS spectra, we considered the
0.7-10.0 keV range.  Above and below this range, strong deviations
from reasonable models are found, indicative of calibration uncertainties.
In our analysis of the HXD/pin spectrum, we considered the 12.0--30.0
keV range.  Fairall 9 is not clearly detected above 30~keV.

\centerline{\psfig{file=f4.ps,width=3.3in,angle=-90}}
\figcaption[h]{This panel above shows the spectra of Fairall 9 fit
with the ``reflionx'' model and the resultant data/model ratio. This
model assumes an ionized disk, but with more ionization species than
the CDID model.  This model provides the best overall fit to the
spectrum of Fairall 9, and suggests a spin of $a = 0.60\pm 0.07$.  The
XIS0, XIS1, XIS3, and HXD spectra are shown in black, red, green, and
blue, respectively.}
\medskip

\section{Analysis and Results}
We used XSPEC v.11.3 (Arnaud 1996) for all of the spectral analysis
reported in this work.  All errors stated were found using the
``error'' and ``steppar'' commands in XSPEC, and correspond to the
1$\sigma$ level of confidence.  The energy of all spectral lines is
reported in the source frame, unless otherwise noted.  All of the
spectral models described below were modified by an interstellar
absorption column density set to $3\times 10^{20}~{\rm cm}^{-2}$
(Dickey \& Lockman 1990) via the ``phabs'' model.  In all fits, we
tied the parameters for all four cameras together but allowed a
constant factor to float between the HXD and XIS detectors to account
for absolute flux offsets.  Our best-fit model (see below) gives a
value of 1.16 for the normalizing constant, consistent with current
calibrations (see, e.g., {\it Suzaku} Memo 2008-06).

We first fit the data with a power-law between 2--3 keV and 7--10 keV.
The data/model ratio resulting from this initial fit is shown in
Figure 1.  This exercise revealed several characteristic spectral
features.  Below 2~keV, there is a soft flux excess that has sometimes
been modeled using a kT$=0.2$~keV blackbody or disk blackbody.  Where
required, we fit this component with a disk blackbody in the models
described below, but we caution that this is a fiducial model to
account for the flux, not strong evidence of such a hot disk (see
Crummy et al.\ 2006 for a more physical treatment of the soft excess).
Above 9 keV, there is a weak excess above the power-law, particularly
in the HXD data. This excess is consistent with a
Compton-backscattering hump due to disk reflection of the incident
hard X-ray flux.

The putative hard flux excess is consistent with the presence of Fe K
emission lines, which also arise through reflection.  The most
prominent emission line is narrow and has a measured energy of
6.10~keV, or 6.40~keV in the frame of Fairall 9.  Narrow Fe K$\alpha$
lines are common in the X-ray spectra of AGN, and may result from
illumination of the ``torus'' (Nandra 2006).  In all models
discussed below, we fit this line with a simple Gaussian function of
zero width, since the line is not resolved.  For consistency, we also
fit an Fe~K$\beta$ line at the proper energy, and with its flux
constrained to be 0.16 times that of the K$\alpha$ line
(Molendi, Bianchi, \& Matt 2003).  After these narrow lines are fit, a broad
asymmetric line is revealed, consistent with reflection from the inner
accretion disk (see Figure 3).

As simple models are easily reproducible, we adopted a 

\centerline{\psfig{file=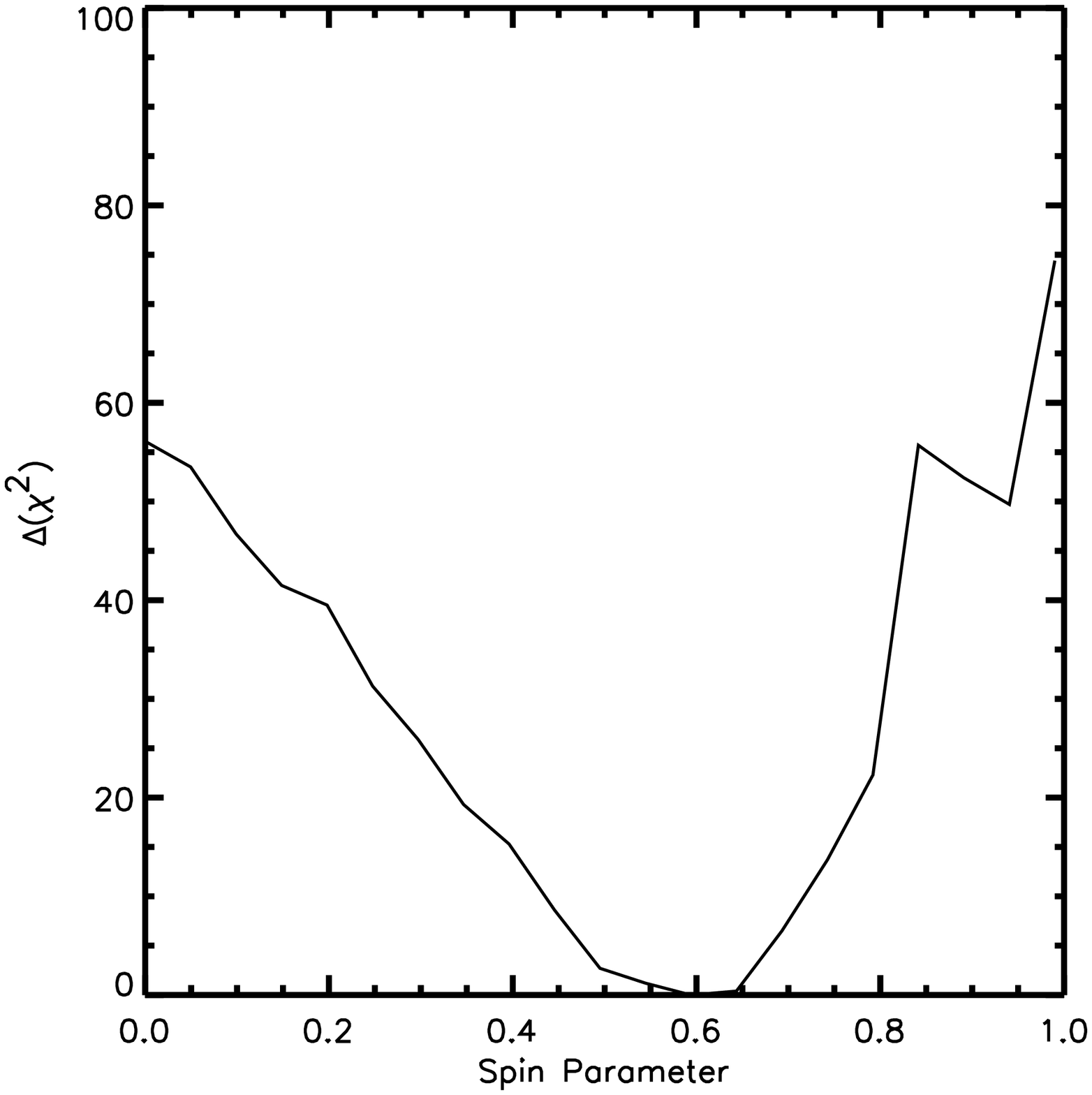,width=3.3in}}
\figcaption[h]{This panel above shows how the goodness-of-fit
statistic varies with black hole spin, when the spectra of Fairall 9
are fit with the ``reflionx'' disk reflection model.  Extremal values
are clearly excluded by the data.}
\medskip

\noindent simple disk
blackbody plus power-law model in order to measure a flux.  We measure
a 0.5--10.0 keV flux of $4.0(1) \times 10^{-11}~ {\rm erg}~{\rm
cm}^{-2}~{\rm s}^{-1}$, and a 0.5--30.0 keV flux of $5.9(2) \times
10^{-11}~ {\rm erg}~{\rm cm}^{-2}~{\rm s}^{-1}$.  The latter flux
corresponds to an X-ray luminosity of $2.6(1)\times 10^{44}~ {\rm
erg}~{\rm s}^{-1}$.  With the bolometric correction found by Marconi
\& Hunt (2003), this corresponds to an Eddington fraction of
approximately 0.13.

With evidence for disk reflection both in the Fe K band and in hard
X-rays, we proceeded to fit the spectrum with common reflection
models.  The reflection models were convolved with a relativistic line
function to account for the relativistic Doppler and gravitational
red-shifts expected near to the black hole (Brenneman \& Reynolds
2006).  With the assumption that the accretion disk is truncated at
the ISCO (consistent with the high Eddington fraction of Fairall 9;
also see Miller et al.\ 2006), the degree to which the line and
reflection spectrum are skewed can be used to constrain the spin of
the black hole in Fairall 9.  These fits are discussed in detail in
the following section.

\subsection{Neutral Disk Reflection}

The ``pexrav'' model describes the reflection of an exponentially
cut-off power-law spectrum from a neutral disk (Magdziarz \& Zdziarski
1995).  As noted above, we blurred this spectrum using the
``kerrconv'' model.  Pexrav does not include an emission line, so our
spectral model also included the ``kerrdisk'' line model.  Parameters
common to the line and convolution blurring function were linked for
self-consistency.  A fiducial disk blackbody component (``diskbb'')
was included in the model to account for the soft excess seen in
Figure 1.

Pexrav requires the metal abundance in solar units and the iron
abundance relative to the metal abundance, as fit parameters.  In the
absence of observational constraints on elemental abundances in
Fairall 9, we fixed the metal abundance to 1.0 and ran the fit three
separate times with an iron abundance of 0.5, 1.0, and 2.0 which gave
$\chi^2$ values of 6575.3, 6553.7, 6626.3 (respectively) for 5970
degrees of freedom.  

Since the relativistic line and disk reflection models are necessarily
separate when ``pexrav'' is employed, the equivalent width of the
relativistic line can be measured directly.  We find a line equivalent
width of $W = 130\pm 10$~eV.  Each fit made with pexrav found Fairall
9 to be consistent with low or moderate spin values.  Abundances of
0.5 and 1.0 gave spin values of $a = 0.0^{+0.2}$ and $ a =
0.1_{-0.1}^{+0.5}$, respectively.  In all cases, maximal spin is
excluded at more than the $5\sigma$ level of confidence. 

Since the convolution model has the inner disk inclination as a
variable parameter, while pexrav uses the cosine of that angle, it was
not possible to link these two parameters directly.  The results
discussed above are based on an inclination
of 40 degrees.  This value was selected after fitting the
model with several different inclination values and tracing the
evolution of the goodness-of-fit statistic, and it is in broad
agreement with the inclination found using other models (see below).

The results we obtained with this model are not entirely
self-consistent.  A relativistic line centroid energy of
$6.70^{+0.01}_{-0.03}$~keV is measured, consistent with He-like Fe
XXV.  This is at odds with the assumption of a completely neutral
accretion disk.  Moreover, a reflection fraction of 2.0 is required in
all fits.  Yet a reflection fraction of $\simeq0.7$ is suggested by
the equivalent width of the relativistic line (George \& Fabian 1991).

\subsection{Ionized Disk Reflection}
Our best fits to the data were obtained using ``reflionx'' (Ross \&
Fabian 2005).  This model includes an Fe K emission line, a broad
range of ionization species, and allows the iron abundance to be a
free parameter.  Unlike pexrav, it is an angle-averaged model; the
inclination angle is not a variable paramter in spectral fits.  Owing
to the fact that reflionx includes low energy emission lines that can
be blurred into a pseudocontinuum that could be the origin of the soft
excess (e.g. Crummy et al.\ 2006), we did not include a disk blackbody
when fitting with reflionx.  It should be noted that reflionx requires
a photon power-law index as an input but requires a separate power-law
component to fit the continuum.  Acordingly, we linked the power-law
index between the two components.  Acceptable fits to Fairall 9 could
not be obtained when the Fe abundance and line emissivity parameters
were frozen at their nominal values.  These parameters were therefore
allowed to vary.

With 5970 degrees of freedom, a blurred reflionx model gave a good
fit: $\chi^{2} = 6498.6$ (see Table 1 and Figure 4).  This is
significantly better than the fits achieved with the pexrav.  Whereas
fits with ``pexrav'' required an ionized line despite the assumption
of neutral reflection, ``reflionx'' returns more self-consistent and
reasonable parameter values.  A steep line emissivity is required (a
maximum of $q=5$ was fixed as per the case of light bending near to a
spinning black hole Miniutti et al.\ 2003).  Using reflionx, we
measure a spin of $a = 0.60\pm 0.07$.  A maximal spin is ruled out at
the $10\sigma$ level of confidence, and zero spin is ruled out at more
than the 7$\sigma$ level of confidence (see Figure 5).

Owing to the fact that CCDs are made of Si, effective area curves
often change abruptly in the Si band.  This makes it difficult to
calibrate detector responses in the Si range.  A formally acceptable
fit with reflionx is not found only due to lingering difficulties in
the calibration of the XIS response in the Si band.  In Figure 1,
clear residuals are seen in this narrow band that differ between the
XIS cameras.  The residuals do not affect the broad-band fit
parameters apart from the goodness-of-fit statistic.

To understand the influence of the soft X-ray band on the spin
constraint made with ``reflionx'', we ignored the spectra below 2 keV
and performed new error scans on the spin 

\centerline{\psfig{file=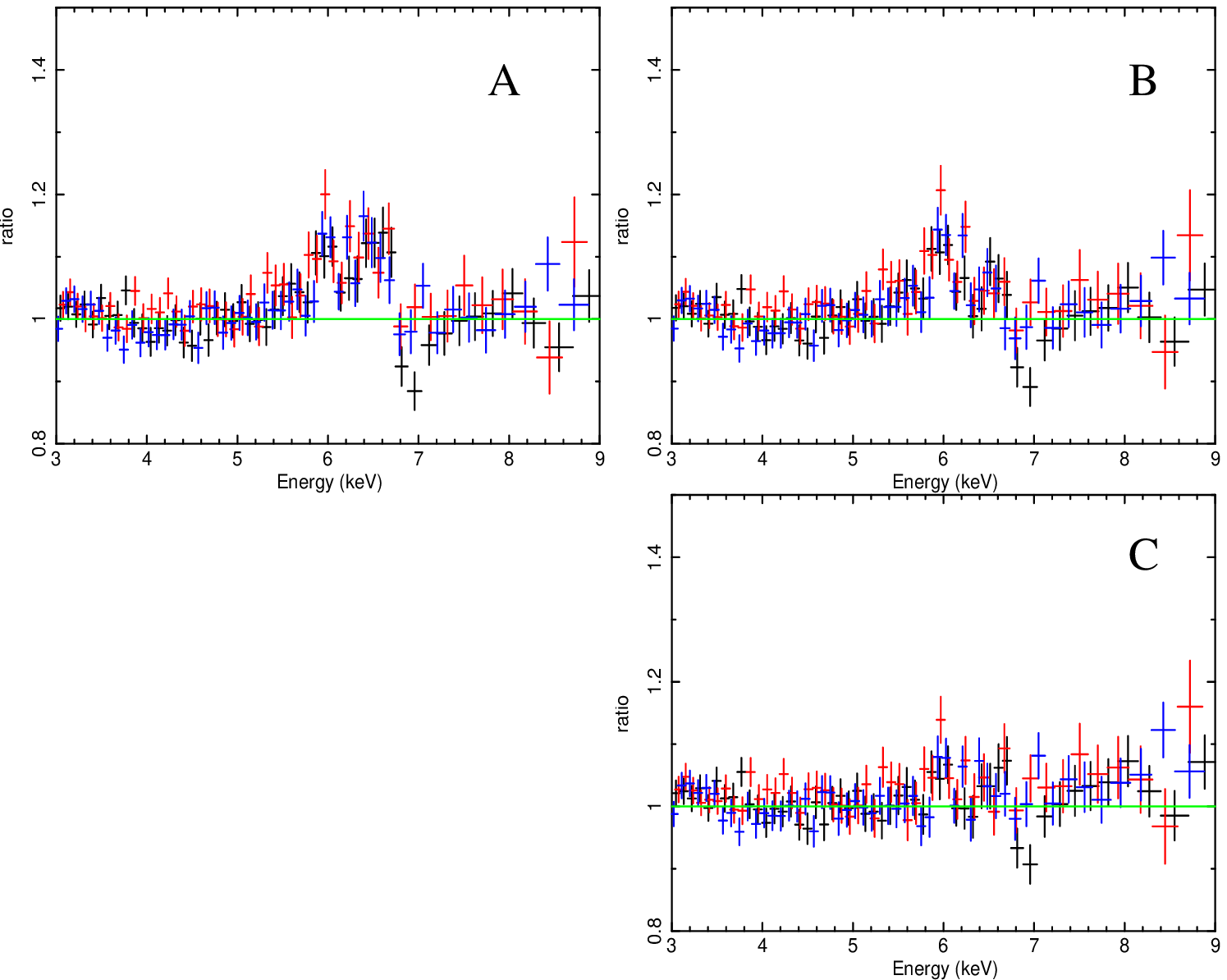,width=3.3in}}
\figcaption[h]{The data/model ratios above show the XIS spectra in the
Fe K band (XIS0 in black, XIS1 in red, XIS3 in blue), after fitting
the models described in Section 3.3.  Panel A shows the data/model
ratio in the Fe K band after two neutral Gaussian lines are added to a
reflection continuum.  Panel B shows the same ratio after two
additional Gaussians, corresponding to Fe XXV and XXVI, have been
added.  Panel C shows the residuals when a single diskline is added
instead of two ionized Fe K lines.}
\medskip

\noindent parameter.  These fits
achieve a significantly worse spin constraint: $a =
0.5^{+0.1}_{-0.3}$.  Zero spin is only excluded at the 90\% level of
confidence; however, maximal spin is excluded at more than the
$5\sigma$ level.\\

\subsection{Inner Reflections}

Though it is common to attribute the spectral features we have
observed to X-ray reflection from the inner disk, it is important to
rigorously rule out alternatives.  This is especially important in
cases like Fairall 9: although the continuum is arguably ``simpler''
than that of Seyfert 1 AGN that have an X-ray warm absorber (see,
e.g. Blustin et al.\ 2005), the signal to noise ratio in the Fe K band
is lower than in cases like MCG-6-30-15 (e.g. Miniutti et al.\ 2007).
We conducted additional investigations to evaluate the
robustness of the disk relection spectrum and relativistic line
interpretation.

The spectra shown in Figures 1, 2, and 3 -- and evidence for a
relativistic disk line -- could be biased by the influence of a disk
reflection spectrum.  We therefore replaced the power-law model used
in those figures with a ``pexrav'' reflection model.  The reflection
fraction was set to 0.7, which corresponds to the equivalent width of
the neutral Fe K line at 6.4~keV (${\rm R} = {\rm EW}/180$~eV; George
\& Fabian 1991).  Note that this is itself conservative, since the
reflection edge is a sharp neutral edge, not one that is broadened by
Compton scattering, and the reflection spectrum was not blurred.  Two
neutral Gaussians corresponding to neutral Fe~K$\alpha$ and K$\beta$
lines were added as before.  A poor fit is achieved with this model
($\chi^{2}/\nu = 6747.4/5975$); broad residuals are still visible in
the spectrum (see Figure 6).

Next, two additional narrow Gaussians corresponding to He-like Fe XXV
and H-like Fe XXVI were included in the model.  These lines were
allowed to vary freely in flux.  This is a very conservative measure:
narrow lines corresponding to Fe XXV or Fe XXVI are extremely rare in
surveys of Seyfert-1 X-ray spectra (see, e.g. Nandra 2007).  Although
this model achieved an improved fit ($\chi^{2}/\nu = 6689.7/5973$),
it still failed to account for all of the flux in the Fe K band (see
Figure 6).  Brenneman \& Reynolds have recently reported evidence for
narrow ionized lines in an {\it XMM-Newton} spectrum of Fairall 9;
however, the lines are single-bin features in a snapshot observation
spectrum.  There is no compelling evidence for distinct lines in our
{\it Suzaku} spectra.

When the two Gaussians corresponding to narrow ionized Fe K lines are
replaced with a single Laor line function with reasonable parameters
(${\rm R}_{\rm in} = 6$~GM/c$^{2}$, $i = 40^{\circ}$, $q = 3$), an
significantly improved fit is achieved ($\chi^{2}/\nu = 6628.7/5973$).
Moreover, the spectrum through the Fe K band is fit well.  An F-test
assuming a difference of one degree of freedom suggests that the
relativistic line model is an improvement over the ionized narrow
Gaussians model at the $7\sigma$ level of confidence.  Compared to
just the neutral Gaussians model, the relativistic disk line model
represents an improvement at more than the 10$\sigma$ level of
confidence (see Figure 6).

\section{Discussion and Conclusions}
We obtained a deep exposure of the Seyfert AGN Fairall 9 using {\it
Suzaku}.  Owing to the sensitivity of the HXD, the source was detected
out to 30~keV for the first time.  The high energy spectrum is
consistent with disk reflection, commensurate with the broad Fe K line
detected in the XIS spectra.  We therefore fit the spectra with two
different disk reflection spectra, each convolved with a relativistic
line function in which the spin of the black hole is a free parameter.
With the assumption that the inner edge of the accretion disk extends
to the ISCO, we are thus able to constrain the spin of the black hole.
Our best fit to the spectrum of Fairall 9 using the ``reflionx'' model
suggests $a = 0.60\pm 0.07$ and excludes extremal spins at a high
level of statistical confidence.  

We employed a number of steps and tests for the purpose of evaluating
the quality of the reflection fit and the robustness of the spin
constraint:\\
$\bullet$ A model including multiple narrow Gaussians from different charge
states of iron, even when added to a raw reflection spectrum, achieves
a fit that is significantly worse than models with a relativistic line.\\
$\bullet$ Fits with a model for reflection from a neutral disk (``pexrav'')
provided a significantly worse description of the data, but yield spin
constraints commensurate with our best-fit model.  Pexrav includes a
sharp neutral Fe K edge, whereas the spectra appear to require ionized
reflection (and a Comptonized edge).  Although fits with ``pexrav''
marked an important test, the model is not ideally suited to the observed
spectra.\\  
$\bullet$ Excluding bins below 2~keV, our best-fit
relativistically-blurred ionized disk reflection model (``reflionx'')
yielded a spin constraint commensurate with that achieved through fits
to the full spectral band, though with larger errors.  The larger
errors are likely the result of two influences: lower overall
sensitivity owing to the exclusion of the band with the greatest
number of photons, and an inability to detect the soft excess (which
may be a psuedocontinuum of blurred disk emission lines; see Crummy et
al.\ 2006).

A broad Fe K line was previously detected in Fairall 9 using ASCA
(Reynolds et al.\ 1997).  More recently, a short {\it XMM-Newton}
observation did not achieve the sensitivity required to detect such a
feature (Gondoin et al.\ 2001; also see Brenneman \& Reynold 2009).  A
broad Fe K line is not only detected in our {\it Suzaku} spectra, but
was easily separated from distinct narrow lines and shown to be
asymmetric and therefore consistent with a disk origin.  This was
possible because of the resolution of the XIS CCDs and the sensitivity
achieved in our long exposure.  Of the sources wherein relativistic
disk lines have been detected (see, e.g., Miller 2007 and Nandra et
al.\ 2007), Fairall 9 is among the most massive (Peterson et
al. 2004).

Our results add to a preliminary picture suggesting that there may be
a range of black hole spin parameters even within a rather homogeneous
class of AGN.  Brenneman \& Reynolds (2006) have reported a spin
approaching $a \simeq 0.99$ in the Seyfert-1 AGN MCG--6-30-15
(Reynolds \& Fabian 2008 suggest that systematics could bring the spin
down to $a = 0.92$), and Miniutti et al.\ (2009) have reported a spin
parameter of $a = 0.6\pm 0.2$ in the narrow-line Seyfert-1 AGN SWIFT
J2127.4$+$5654.  These results are self-consistent in that they all
derive from spectral fits with relativistically-blurred disk
reflection models.  A similar picture may be emerging from studies of
stellar-mass black holes (Miller et al.\ 2009).

In each of these three cases, a steep line emissivity ($q\simeq 5$) is
required.  This value is consistent with theoretical predictions of
the influence of gravitational light bending near to a spinning black
hole (Miniutti et al.\ 2003).  However, MCG-6-30-15 is the only case
among the three wherein a high spin parameter has been inferred.  The
proper form of the the emissivity index is largely unknown, at
present.  In the case of Fairall 9, the emissivity index could be as
high as $q=6$, though an upper bound of $q=5$ is enforced in our fits.
(The spin constraints are consistent for $q=5$ and $q=6$).
Observations of microlensing in quasars indicates that the hard X-ray
emission region is smaller than $6~GM/c^{2}$ (Chartas et al.\ 2009).
Such results bolster implications for light bending and steep
line emissivity parameters.

A range of spin parameters is allowed by current theoretical
investigations.  Based on calculations performed by Berti \& Volonteri
(2008) that take into consideration both black hole mergers and
accretion onto black holes, the probability of a spin as high as that
of MCG--6-30-15 is less than $10^{-4}$ in the chaotic accretion model
by King et al. (2008).  Alternatively, if accretion occurs in a
coherent fashion, the spin distribution should peak at high value
(e.g. $a \geq 0.9$).  However, the predicted distributions have a
small tail extending to zero spin: the probability of a spin $<0.1$ in
the coherent accretion scenario is 10\%.  Volonteri et al. (2007) also
discuss how the distribution of spins in low-redshift AGN is likely
broader than the theoretical limit derived for quasars in the coherent
accretion model, which requires galaxy mergers to drive large gas
inflows and to trigger quasar activity.  Thus, the fact that our
models imply that the black hole in Fairall 9 may have a moderate spin
is consistent with current theories.

However, with only three spin constraints, we cannot yet rule out any
of the theoretical models, despite their striking differences.  Many
more spin constraints are needed to distinguish between different
predictions of the cosmological black hole spin distribution
(Volonteri et al.\ 2005, King \& Pringle 2006).  Our results
demonstrate that the sensitivity and resolution of the XIS cameras,
and the sensitivity and low background of the HXD, are extremely
helpful in measuring disk reflection spectra.  Over the lifetime of
the {\it Suzaku} mission, spin constraints may be possible in as many
as 20 Seyfert AGN.  In the long run, the much larger collecting area
of the {\it International X-ray Observatory} will make it possible to
obtain spin measurements in as many as 300 AGN (Brenneman et al.\
2009).

\vspace{0.2in}

We thank the anonymous referee for helpful comments that improved this
  work.  J. M. M. gratefully acknowledges funding from NASA through
  the {\it Suzaku} guest investigator program.  E. M. C. gratefully
  acknowledges support provided by NASA through the {\it Chandra}
  Fellowship Program, grant number PF8-90052.  We thank the US and
  Japanese {\it Suzaku} teams for executing this observation.  We
  acknowledge helpful discussions with Koji Mukai and Oleg Gnedin.
  This work has made us of the facilities and tools available through
  HEASARC, operated by GSFC for NASA.

\begin{table}[t]
\begin{center}
\caption{Summary of Model Parameters}
\vspace{0.1in}
\begin{footnotesize}
\begin{tabular}{ll}
\hline
{\bf Parameter} & {\bf Reflionx} \\ 
\hline\\
Fe K-$\alpha$ narrow line (keV) & $6.400_{-0.002}^{+0.002}$ \\
Fe K-$\alpha$ narrow line norm ($10^{-5}$) & $2.2^{+0.2}_{-0.2}$ \\
Spin ($cJ/GM^{2}$) & $0.65^{+0.05}_{-0.05}$\\
Inclination (deg) & $44^{+1}_{-1}$\\
Emissivity Index & $5.0_{-0.1}$\\
Inner Radius ($R_{ISCO}$) & 1.0*\\
Outer Radius ($R_{ISCO}$) & 400.0* \\
$\xi$ (erg cm ${\rm s}^{-1}$) & $3.7^{+0.1}_{-0.1}$ \\
Photon Index & $2.09^{+0.01}_{-0.01}$\\
Fe Abundance & $0.8^{+0.2}_{-0.1}$  \\
Reflection norm ($10^{-5}$) & $9^{+1}_{-1}$\\ \\ \hline
chi-squared & 6503.7\\   
dof & 5969\\  \hline \hline
\end{tabular}
\vspace{-0.2in}
\tablecomments{The parameters of our best-fit relativistically-blurred
disk reflection model are given above.  Frozen parameters are marked
with an asterisk.  The normalization of the power-law component
external to the ``reflionx'' model is $8.3(1)\times 10^{-3}$.  A hard
limit of $q=5$ was set on the line emissivity index as per Merloni \&
Fabian (2003).  The inner and outer blurring radii correspond to the
innermost stable circular orbit (ISCO) for the given spin parameter.\\
}
\end{footnotesize}
\end{center}
\end{table}

\end{document}